\documentclass[amsmath,amssymb,prb,superscriptaddress,twocolumn,showpacs]{revtex4-1}
\usepackage{graphicx}
\usepackage{dcolumn}
\usepackage[colorlinks=true,linkcolor=blue,citecolor=blue,urlcolor=blue]{hyperref}

\newcommand{\HH}{\mathcal{H}}
\newcommand{\ud}{\mathrm{d}}
\newcommand{\iu}{\mathrm{i}}
\newcommand{\Tr}{\mathrm{Tr}}
\newcommand{\IM}{\mathrm{Im}}

\newcommand{\muB}{\mu_{\text{B}}}
\newcommand{\upup}{\uparrow\uparrow}
\newcommand{\dndn}{\downarrow\downarrow}

\begin{document}

\title{Relativistic dynamical spin excitations of magnetic adatoms}

\author{M. dos Santos Dias}\email{m.dos.santos.dias@fz-juelich.de}
\author{B. Schweflinghaus}
\author{S. Bl\"ugel}
\author{S. Lounis}\email{s.lounis@fz-juelich.de}
\affiliation{Peter Gr\"{u}nberg Institut and Institute for Advanced Simulation, Forschungszentrum J\"{u}lich \& JARA, D-52428 J\"{u}lich, Germany}

\date{\today}

\begin{abstract}
\noindent
We present a first-principles theory of dynamical spin excitations in the presence of spin-orbit coupling.
The broken global spin rotational invariance leads to a new sum rule.
We explore the competition between the magnetic anisotropy energy and the external magnetic field, as well as the role of electron-hole excitations, through calculations for 3$d$-metal adatoms on the Cu(111) surface.
The spin excitation resonance energy and lifetime display non-trivial behavior, establishing the strong impact of relativistic effects.
We legitimate the use of the Landau-Lifshitz-Gilbert equation down to the atomic limit, but with parameters that differ from a stationary theory.
\end{abstract}



\maketitle

\section{Introduction}\label{sec:intro}

The understanding and design of new technologies based on magnetic materials, in the fields of spintronics and magnonics, begets and profits from quantitative 
theoretical approaches.
Recently, the central role played by spin-orbit coupling (SOC) has been serendipitously revealed.
It underlies many concepts for the generation, manipulation and detection of spin currents~\cite{Maekawa2012}, also at the nanoscale: spin dynamics, current-induced magnetization switching and magnetic stability~\cite{Loth2010,Khajetoorians2013,Oberg2014}.
The energy gap in the spin excitation spectrum, from extended materials down to a single adatom, is the result of SOC, being measurable, for instance, with ferromagnetic resonance (FMR)~\cite{Farle1998} or inelastic scanning tunneling spectroscopy (ISTS)~\cite{Heinrich2004,Hirjibehedin2006,Khajetoorians2011,Chilian2011,Khajetoorians2013a,Balashov2009}.
Physical intuition suggests a connection between the gap and the magnetic anisotropy energy (MAE), while application of a dc external magnetic field ($B^{\text{ext}}$) should lead to a gap change of $g \muB B^{\text{ext}}$ (Zeeman shift).
Stoner (electron-hole) excitations, together with SOC, contribute to the spin dynamics, their impact on the energy gap, Zeeman shift and excitation lifetime being expected but fairly unexplored.

Up to now, all \emph{ab initio} studies on transverse dynamical spin excitations have neglected SOC.
Furthermore, the addressed magnetic states are collinear (\emph{e.g.}~ferromagnetic), since the complexity, both theoretical and computational, increases dramatically in the general case: charge excitations, longitudinal and transverse spin excitations may couple in a non-trivial manner (see discussion in Ref.~\onlinecite{Kambersky1985}). 
From the tight-binding perspective, the work of Costa {\it et al.}~\cite{Costa2010} has proven invaluable in characterizing the impact of SOC on the spin-wave dispersion and lifetime, going beyond the adiabatic approximation~\cite{Udvardi2009}. 
Two kinds of theoretical approaches build upon first-principles electronic structure calculations.
One type is based on many-body perturbation theory (MBPT)~\cite{Karlsson2000,Kotani2008,Sasioglu2010}, constructing the non-interacting Green function (GF) from the density functional theory (DFT) eigenstates. 
The other type is based on time-dependent DFT (TDDFT)~\cite{Gross1985a,Liu1989}, and has been applied to bulk systems~\cite{Savrasov1998,Zhukov2004,Kotani2008,Buczek2009,Sasioglu2010,Rousseau2012}, thin films~\cite{Buczek2011} and adatoms on surfaces~\cite{Lounis2010,Lounis2011,Lounis2014}, with pioneering work on SOC~\cite{Matsumoto1990}, but still not starting from a spin-polarized ground state~\footnote{TD spin-current DFT is an alternative~\cite{Vignale1987,Vignale1988,Bencheikh2003}.}.
Recently we provided a connection between MBPT and TDDFT, to describe the interaction between electrons and spin excitations~\cite{Schweflinghaus2014}; this link was also found  for the theory of spin-fluctuation-mediated superconductivity~\cite{Essenberger2014}.

Here we present a method for the calculation of dynamical magnetic response functions based on TDDFT and incorporating SOC.
This new scheme is implemented within the Korringa-Kohn-Rostoker (KKR) GF approach~\cite{Papanikolaou2002}.
Of interest is the ability to treat an external magnetic field ($\vec{B}^{\mathrm{ext}}$) in any direction, to investigate non-trivial orientations of the magnetic moments and the role of the MAE.
We derive a novel magnetization sum rule that constrains the exchange and correlation (xc) kernel when SOC is present, which is essential for the theory and the calculations.
By investigating $3d$ metal adatoms on the Cu(111) surface, with a focus on the experimentally studied Fe adatom~\cite{Khajetoorians2011}, we demonstrate that the gap in the excitation spectrum is connected to the MAE through spin dynamics parameters, with electron-hole excitations taking center stage.
We also show that, varying the orientation of $\vec{B}^{\mathrm{ext}}$, the properties of the spin-excitations, \emph{i.e.}~excitation energies, lifetimes and $g$-factors, are non-trivial and anisotropic, depending on the orientation of the magnetic moment.

Our paper is organized as follows.
In Sec.~\ref{sec:dynsusc} the formalism for the dynamical magnetic susceptibility is presented, as is the new magnetization sum rule that applies for the SOC case.
Sec.~\ref{sec:compdet} describes the computational details.
The results for 3d adatoms on Cu(111) are discussed in Sec.~\ref{sec:alladatoms}, and the interplay between MAE and SOC is explained in Sec.~\ref{sec:feadatom}.
Our conclusions are given in Sec.~\ref{sec:conc}, while three appendices provide further details.

\section{Dynamical susceptibility}\label{sec:dynsusc}

We begin by seeking the change of the spin density matrix, $\delta\rho = \delta n\,\sigma_0 + \delta\vec{m}\cdot\vec{\sigma}$, due to an external time-dependent potential of the same form, $\delta V^{\text{ext}} = \delta V_0^{\text{ext}}\,\sigma_0 + \delta\vec{B}^{\text{ext}}\cdot\vec{\sigma}$.
In linear response,
\begin{equation}
  \delta\rho(\vec{r}\,;t) = \!\int\!\ud\vec{r}\,'\!\!\int\!\ud t'\,\chi(\vec{r}\,,\vec{r}\,';t-t')\cdot\delta V^{\text{ext}}(\vec{r}\,';t') \;\;,
\end{equation}
with $\vec{\sigma}$ the vector of Pauli matrices and $\sigma_0$ the $2\times2$ unit matrix.
Within TDDFT, after Fourier transforming in time, the full response function, $\chi$, is given in terms of the Kohn-Sham (KS) response, $\chi^{\text{KS}}$, and the Hartree-exchange-correlation (Hxc) kernel, $K^{\text{Hxc}}$, through
\begin{equation}
  \chi(\omega) =  \left[1- \chi^{\text{KS}}(\omega)\,K^{\text{Hxc}}(\omega)\right]^{-1} \chi^{\text{KS}}(\omega)\label{suscdyson}
\end{equation}
(here and in the following spatial dependence and integrations are omitted for brevity).
We add SOC~\footnote{The radial SOC potential is $\xi(r) = \frac{1}{(Mc)^2}\frac{1}{r}\frac{\ud V}{\ud r}$} and spin and orbital Zeeman couplings to an external static field to the KS Hamiltonian, self-consistently,
\begin{equation}
  \HH^{\text{KS}} = \HH^{\text{KS}}_0\sigma_0 + \vec{\sigma}\cdot\vec{B}^{\text{xc}} + \xi\,\vec{L}\cdot\vec{\sigma} + \big(\vec{L} + \vec{\sigma}\big)\cdot\vec{B}^{\text{ext}} \;\;.
  \label{eq:soczeeman}
\end{equation}
The KS response is written in terms of the KS GF, $G(E) = \big(E - \HH^{\text{KS}}\big)^{\text{--}1}$, as
\begin{align}
  \hspace{-0.5em}\chi^{\text{KS}}_{\alpha\beta}(\omega) = -\frac{1}{\pi}\!\int^{E_{\text{F}}}\!\!\!\!\!\!\!\ud&E\;\Tr \big[ \sigma_\alpha\,G(E + \omega + \iu0)\,\sigma_\beta\,\IM\,G(E) \nonumber\\
  &+ \sigma_\alpha\,\IM\,G(E)\,\sigma_\beta\,G(E-\omega-\iu0) \big] \;, \label{kssusc}
\end{align}
with $\alpha,\beta = x,y,z$.
The trace is over the spin components, and $2\iu\,\IM\,G(E) = G(E+\iu0) - G(E-\iu0)$.
$K^{\text{Hxc}}(\omega)$ is the sum of the Hartree kernel, $2/|\vec{r} - \vec{r}\,'|$, and of the xc kernel, which contains the many-body effects.
In the adiabatic local spin density approximation (ALSDA), adopted in this work, the xc kernel is local in space and frequency-independent, being a function(al) of the particle and spin densities only.

In the absence of SOC, the full response function decouples into a transverse and a longitudinal part.
The former describes damped precessional motions of the magnetization, while excitations that change the magnitude of the charge or spin densities are in the latter.
Global SU(2) invariance implies the existence of a Goldstone mode: the spin density is an eigenfunction of $\chi^{-1}(\omega\!=\!0)$ with vanishing eigenvalue.
This is of utmost importance in numerical calculations, as small inaccuracies in the KS susceptibility and in the xc kernel shift the Goldstone mode to a finite frequency, in the meV range~\cite{Lounis2010,Lounis2011}, where the gap opened by SOC is expected. 
Corrective schemes were proposed~\cite{Lounis2010,Lounis2011,Rousseau2012} by adjusting one or both of these quantities to place the Goldstone mode at zero frequency, when SOC is neglected.
Next we derive a new sum rule, connecting the spin density to the xc magnetic field, SOC and external magnetic field, so that the gap arises unambigously from the latter two.

Suppose that the xc magnetic field, $\vec{B}^{\text{xc}}$, lies in the $z$-direction, defining the local spin frame of reference \footnote{If the direction of the xc magnetic field is not the $z$-direction, the final expressions hold after an appropriate rotation of the coordinate system.}. 
Then the spin density is given by (see Appendix~\ref{app:sumrule}):
\begin{align}
  m_z &= -\frac{1}{\pi}\,\IM\,\Tr\!\int^{E_{\text{F}}}\!\!\!\!\!\!\!\ud E\,G_{\upup}(E)\,\Delta(E)\,G_{\dndn}(E) \;\;.\hspace{-0.5em} \label{sumrulederivation}
\end{align}
The effective spin splitting is, \emph{cf.}~Eq.~(\ref{eq:soczeeman}) and Appendix~\ref{app:sumrule}:
\begin{align}
  \Delta &= \HH^{\text{KS}}_{\uparrow\uparrow} - \HH^{\text{KS}}_{\downarrow\downarrow}
 + \HH^{\text{KS}}_{\uparrow\downarrow}\,\widetilde G_\downarrow\,\HH^{\text{KS}}_{\downarrow\uparrow}
 - \HH^{\text{KS}}_{\downarrow\uparrow}\,\widetilde G_\uparrow\,\HH^{\text{KS}}_{\uparrow\downarrow} \label{delta} \;\;,\hspace{-0.5em}
\end{align}
where the auxiliary GFs are
$\widetilde G_\sigma(E) = \big(E - \HH^{\text{KS}}_{\sigma\sigma}\big)^{\text{--}1}$, with $\sigma=\;\uparrow,\downarrow$.
Identifying the spin-flip KS susceptibility via Eq.~(\ref{kssusc}),
$\chi_{+-} = \big(\chi_{xx} - \iu\chi_{xy} + \iu\chi_{yx} + \chi_{yy}\big)/4$, 
we can relate the spin density to the xc magnetic field,
$m_z = 2\,\chi^{\text{KS}}_{+-}(\omega=0)\,B^{\text{xc}} + \delta m_z$,
where $\delta m_z$ arises from all contributions to $\Delta$, excluding the xc part. 
The transverse xc kernel in the ALSDA is just $K_\perp = 2 B^{\text{xc}}/m_z$, yielding the magnetization sum rule
\begin{equation}
  m_z(\vec{r}\,) - \!\int\!\!\ud\vec{r}\,'\,\chi^{\text{KS}}_{+-}(\vec{r}\,,\vec{r}\,';0)\,K_\perp(\vec{r}\,')\,m_z(\vec{r}\,') = \delta m_z(\vec{r}\,) \;\;. \label{sumrule1}
\end{equation}
When $\delta m_z(\vec{r}\,) = 0$ (no SOC or external field), the denominator of Eq.~(\ref{suscdyson}) vanishes and the Goldstone mode is recovered.
If only an external field is applied, the spin excitation is located at $\omega \sim 2B^{\text{ext}}_z$ (Zeeman shift), while SOC gives rise to a gap even in zero field. 
If Eq.~(\ref{sumrule1}) is not satisfied due to numerical inaccuracies, $K_\perp(\vec{r}\,)$ is adjusted such that the sum rule holds, using as input the calculated $\delta m_z(\vec{r}\,)$, $m_z(\vec{r}\,)$ and $\chi^{\text{KS}}_{+-}(\vec{r}\,,\vec{r}\,';\omega=0)$.

Next we briefly summarize the computational details, before presenting results of our TDDFT formalism for adatoms on Cu(111).

\section{Computational details}\label{sec:compdet}

The GFs are evaluated by the KKR method~\cite{Papanikolaou2002} where the electronic structure of the adatoms is computed in two steps.
First the electronic structure of a 22 layer Cu slab is calculated.
Then each adatom is self-consistently embedded on the surface of this slab, in real space, together with 12 nearest-neighbor sites.
We consider the Fe adatom relaxed to the surface by 10\% of the interlayer distance of Cu and use, for the sake of comparison, the same vertical distance for all adatoms.
We employ the rigid spin approximation, whereby the direction of the spin density is taken to be collinear inside each atomic sphere \footnote{Spin-orbit coupling is a source of intra-atomic non-collinearity of the spin density; this is a small effect for the adatoms considered in this study.}.
To tackle canting of the spin moment due to perpendicular external magnetic field and anisotropy easy axis, its direction is updated during the iterations until self-consistency is achieved.
The MAE is calculated by band energy differences following the magnetic force theorem~\cite{Weinert1985},
\begin{equation}
  \quad E_a \approx E_{\text{band}}[B^{\text{xc}}\hat{e}_x] - E_{\text{band}}[B^{\text{xc}}\hat{e}_z] \;\;. \label{mae}
\end{equation}
For the susceptibility calculations, the proposal of Lounis \emph{et al.}~\cite{Lounis2010,Lounis2011} is extended to an $spdf$ basis built out of regular scattering solutions evaluated at two or more energies, by orthogonalizing their overlap matrix (see Appendix~\ref{app:suscbasis} for details).
This basis can reproduce ground state data reliably and is used to include SOC in the GFs, self-consistently
The spatial dependence of the susceptibility is restricted to the magnetic adatom; tests including neighboring copper atoms show no significant impact on the transverse spin excitations.
The energy integration in Eq.~(\ref{kssusc}) is performed as detailed in Refs.~\onlinecite{Lounis2010,Lounis2011}.

\section{Spin excitations of C\lowercase{r}, M\lowercase{n}, F\lowercase{e} and C\lowercase{o} adatoms on the C\lowercase{u}(111) surface}\label{sec:alladatoms}

As an application of our method, we explore the spin excitation spectra of $3d$ adatoms on the Cu(111) surface.
The groundstate properties and the dynamical spin excitation spectra are described separately.

\subsection{Ground state properties}

\renewcommand{\arraystretch}{1.1}
\begin{table}[b]
\begin{ruledtabular}
  \begin{tabular}{c c c c c}
   & Cr & Mn & Fe & Co \\
  $m^{\text{s}}$ ($\muB$) &  $\phantom{-}4.07$ & 4.31 & 3.23 & 1.97 \\
  $m^{\text{o}}$  ($\muB$) &  $-0.02$           & 0.02 & 0.55 & 0.52 \\
  $E_a$ (meV)                  & $-0.29$ & $-0.33$ & 4.96 & 2.25 \\
  \end{tabular}
\end{ruledtabular}
\caption{\label{tab1}Spin ($m^{\text{s}}$) and orbital ($m^{\text{o}}$) moments for Cr, Mn, Fe and Co adatoms on the Cu(111) surface, and MAE, see Eq.~(\ref{mae}).
All values for the easy axis configuration (in-plane for Cr and Mn, out-of-plane for Fe and Co).}
\end{table}

\begin{figure}[t]
  \includegraphics[width=\columnwidth]{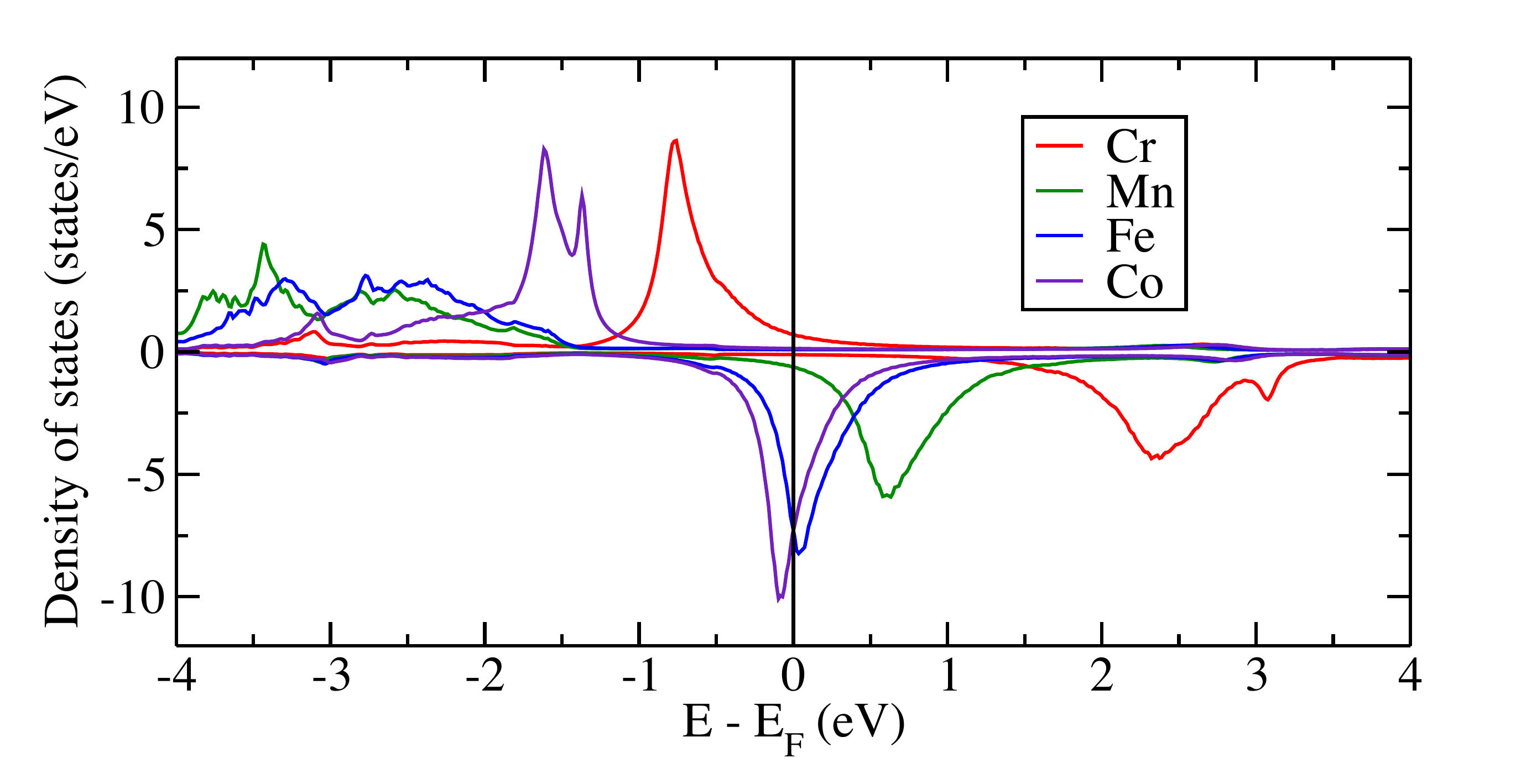}\vspace{-1em}
  \caption{\label{fig1} Atom-projected total density of states for Cr, Mn, Fe and Co adatoms on the Cu(111) surface (positive for majority and negative for minority spin).
  Energy measured from the Fermi energy of the substrate.}
\end{figure}

The adatom-projected local density of states for Cr, Mn, Fe and Co adatoms on the Cu(111) surface is shown in Fig.~\ref{fig1}, while ground state properties are listed in Table~\ref{tab1}.
The spin moment, $m^{\text{s}}$, is maximum for Cr and Mn, and then decreases steadily for Fe and Co.
The orbital moment, $m^{\text{o}}$, is small for Cr and Mn, a consequence of the almost half-filled $d$-states, and large for Fe and Co, due to partial occupation of the minority $d$-states.
As expected, Cr and Mn adatoms with nearly half-filled $d$-shells have larger spin magnetic moments ($m^{\text{s}}$) and lower orbital magnetic moments ($m^{\text{o}}$) than those of Fe and Co adatoms. 
The small magnitude of the orbital moments for Cr and Mn correlates with small MAEs, the converse being true for Fe and Co.
The preferred orientation of the magnetic moments of Fe and Co is normal to the surface while Cr and Mn lie in-plane.

\subsection{Dynamical spin excitations}

\begin{figure}[t]
  \includegraphics[width=\columnwidth]{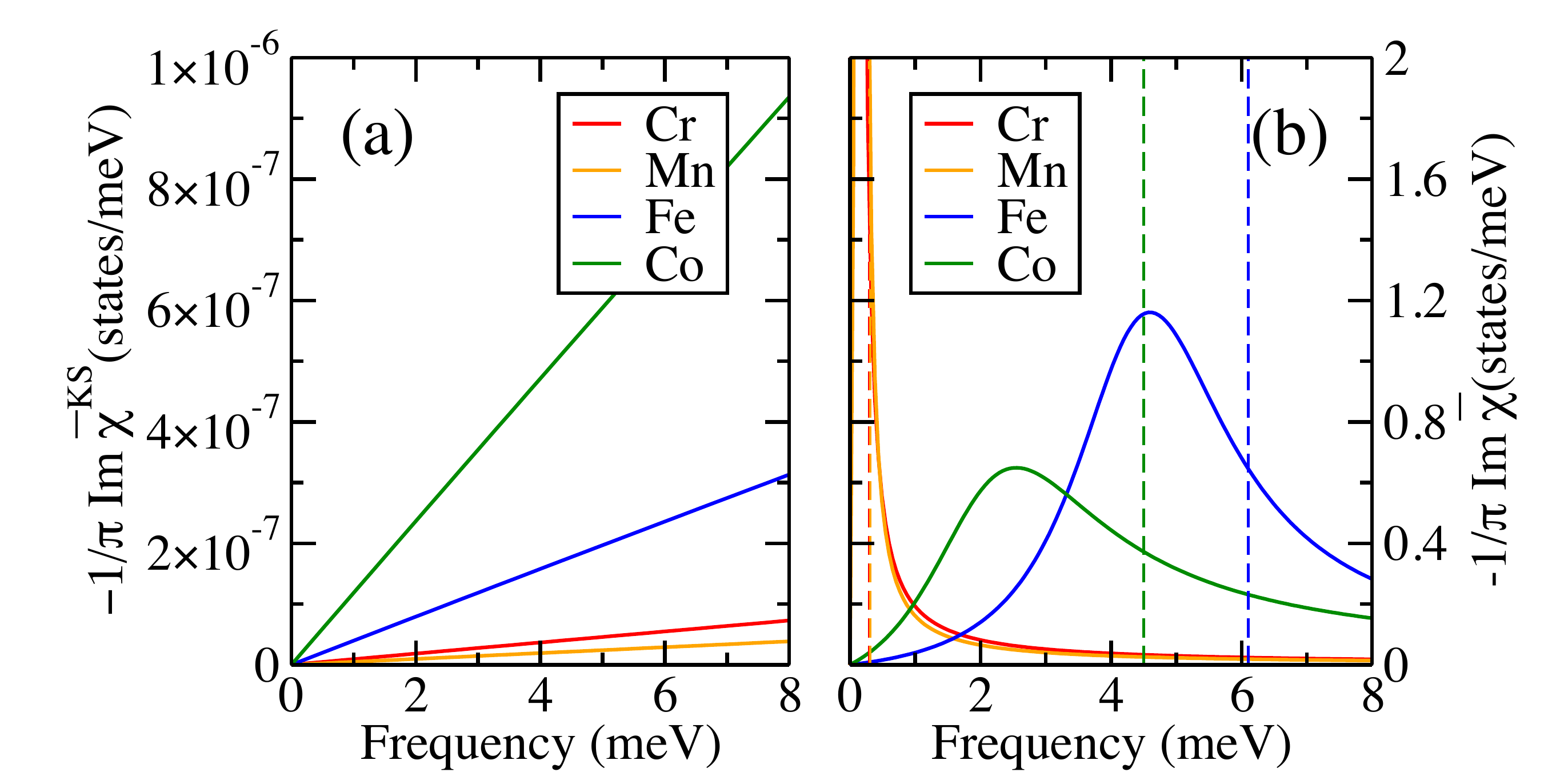}\vspace{-1em}
  \caption{\label{fig2} Density of transverse spin excitations for Cr, Mn, Fe and Co adatoms on the Cu(111) surface ($B^{\text{ext}} = 0$).
  (a) from  $\IM\,\overline{\chi}^{\text{KS}}_{+-}(\omega)$, Eq.~(\ref{kssusc}).
  (b) from  $\IM\,\overline{\chi}_{+-}(\omega)$, Eq.~(\ref{suscdyson}).
  Dashed lines show the result of the estimate based on the adiabatic approximation (see main text).
  }\vspace{-1em}
\end{figure}

Now we proceed to the calculations of the dynamical magnetic susceptibility.
Eq.~(\ref{suscdyson}) is solved using the full basis expansion of the GFs, but for discussion we define an adatom-averaged quantity, corresponding to the net response to a site-dependent TD external magnetic field:
\begin{equation}
  \overline{\chi}_{\alpha\beta}(\omega) = \!\int\!\ud\vec{r}\!\int\!\ud\vec{r}\,'\,\chi_{\alpha\beta}(\vec{r}\,,\vec{r}\,';\omega) \;\;. \label{avgsusc}
\end{equation}
First we discuss the impact of SOC on the adatom-averaged KS susceptibility.
$\overline{\chi}^{\text{KS}}_{+-}(\omega)$ is linear for small frequencies.
Fig.~\ref{fig2}(a) shows $\IM\,\overline{\chi}^{\text{KS}}_{+-}(\omega)$, describing spin-flip excitations between occupied and empty states (Stoner excitations).
SOC is found to have a negligible impact on $\overline{\chi}^{\text{KS}}_{+-}(\omega)$ for the Cr and Mn adatoms, in line with their low orbital magnetic moments and MAEs.
For Fe, SOC causes a noticeable change only on $\overline{\chi}^{\text{KS}}_{+-}(\omega = 0)$ ($\sim\!0.2$\%), while for Co there is also an increase in the slope of $\IM\,\overline{\chi}^{\text{KS}}_{+-}(\omega)$ by 12\%.

Consider now the full response function.
The longitudinal and transverse parts of the magnetic susceptibility are weakly coupled by SOC.
The imaginary part of the dominant eigenvalue of $\overline{\chi}(\omega)$ corresponds to the density of states of the main magnetic excitation, and is shown in Fig.~\ref{fig2}(b) for all adatoms.
Although the resonance energies ($B^{\text{ext}} = 0$) follow the trend of the adatoms' MAE (Table~\ref{tab1}), their positions are strongly shifted from what is expected from a simple approximation (dashed lines in Fig.~\ref{fig2}(b); discussed in the following).

We characterize the main resonance in $\overline{\chi}(\omega)$ through a simple model of the spin dynamics, the Landau-Lifshitz-Gilbert (LLG) equation~\cite{Gilbert2004}, widely used for larger magnetic systems:
\begin{equation}
  \frac{\ud\vec{m}^{\text{s}}}{\ud t} = -\gamma\,\vec{m}^{\text{s}} \times \vec{B}^{\text{eff}} + \eta\,\frac{\vec{m}^{\text{s}}}{m^{\text{s}}}\!\times\!\frac{\ud\vec{m}^{\text{s}}}{\ud t} \;,\;\; \vec{B}^{\text{eff}} = -\frac{\partial E}{\partial\vec{m}^{\text{s}}}  \label{llgeq}
\end{equation}
with $\vec{m}^{\text{s}}$ the spin moment, $m^{\text{s}}$ its length, $\gamma$ the gyromagnetic ratio (equal to $2\muB/\hbar$ for a free electron) and $\eta$ the damping parameter.
We set $\hbar = 1$ and absorb $\muB$ in $\vec{B}^{\text{eff}}$, so $\gamma = 2$.
As discussed by Kittel~\cite{Kittel1949} and found from our calculations, the orbital moment mostly follows the spin moment, $\vec{m}^{\text{o}} \propto \vec{m}^{\text{s}}$, so it is not considered independently.

\renewcommand{\arraystretch}{1.1}
\begin{table}[b]
\begin{ruledtabular}
  \begin{tabular}{c c c c c}
   & Cr & Mn & Fe & Co \\
%  $m_z$ ($\muB$)             &  4.071 &  4.313 & 3.235 & 1.976 \\
%  $B^{\text{eff}}$ (meV)  & 0.071 &  0.010 & 2.79 & 1.39 \\
  $\bar{E}_a$ (meV)     & 0.14 &  0.02 & 4.51 & 1.37 \\
%  $\omega_0$ (meV)        & 0.116 &  0.018 & 4.37 & 2.00 \\
  $\gamma$                &  1.63 &  1.74  & 1.73 & 2.36 \\
  $\eta$                  & 0.07 &  0.05 & 0.33 & 0.80 \\
  $g$ (spin only)         &  1.63 &  1.74  & 1.64 & 1.84 \\
  $g$ (spin+orb)          &  1.62 &  1.75  & 1.92 & 2.33 \\
  $\tau$ (ps)             &  22 &  1200  & 0.11 & 0.091
  \end{tabular}
\end{ruledtabular}
\caption{\label{tab2} Spin dynamics parameters for Cr, Mn, Fe and Co adatoms on the Cu(111) surface, obtained by fitting the transverse susceptibility, Eq.~(\ref{avgsusc}), to the LLG model.
Eq.~(\ref{wmax}) defines $\bar{E}_a$.
The $g$-factor, Eq.~(\ref{gfactor}), distinguishes $\vec{m}^{\text{s}}\cdot\vec{B}^{\text{ext}}$ or $(\vec{m}^{\text{s}} + \vec{m}^{\text{o}})\cdot\vec{B}^{\text{ext}}$  couplings.
$\tau$ for Mn is hard to extract, due to a very narrow spin excitation peak.}
\end{table}

Suppose $\vec{m}^{\text{s}} = m^{\text{s}}\,\hat{e}_z$ in equilibrium, so $\vec{B}^{\text{eff}} = B^{\text{eff}}\,\hat{e}_z$, and we linearize Eq.~(\ref{llgeq}) adding a small time-dependent perturbation in the $xy$-plane.
From $\IM\,\chi_{+-}(\omega)$, the resonance location, $\omega_{\text{max}}$, is given by (Appendix~\ref{app:llg}), 
\begin{equation}
  \omega_{\text{max}} = \frac{\gamma\,B^{\text{eff}}}{\sqrt{1 + \eta^2}} 
  \;,\;\; B^{\text{eff}} = \frac{2{E}_a}{m^{\text{s}}} + \left(1 + \frac{m^{\text{o}}}{m^{\text{s}}}\right) B^{\text{ext}}  \label{wmax}
\end{equation}
with a model energy of the form
\begin{equation}
  E = -\frac{E_a}{(m^{\text{s}})^2}(\vec{m}^{\text{s}}\cdot\hat{e}_z)^2 - (\vec{m}^{\text{s}} + \vec{m}^{\text{o}})\cdot\vec{B}^{\text{ext}} \;\;.
\end{equation}
The spectroscopic $g$-factor is defined as
\begin{equation}
  g = \frac{\ud \omega_{\text{max}}}{\ud B^{\text{ext}}} = \frac{\gamma}{\sqrt{1 + \eta^2}}\left(1 + \frac{m^{\text{o}}}{m^{\text{s}}}\right) \;\;. \label{gfactor}
\end{equation}
A non-zero orbital moment (SOC) implies an effective $g \propto \gamma \left(1+\frac{{m}^{\text{o}}}{{m}^{\text{s}}}\right)$, which is the usual explanation for $g \neq 2$ found experimentally (for instance in FMR~\cite{Farle1998}).
Damping renormalizes $g$ further by a factor $1/\sqrt{1 + \eta^2}\,$, so that a shift can be expected even without SOC (${m}^{\text{o}} = 0$).
Lastly, the inverse of the full-width at half maximum (FWHM, $\Gamma$) of the spin excitation resonance can be used to estimate the corresponding lifetime, $\tau \approx \hbar/(2\Gamma)$.
All model parameters are extracted by fitting the LLG form of $\chi_{+-}(\omega)$ given in Eq.~(\ref{eq:llgimchi}) to the first-principles data in Fig.~\ref{fig2}(b), and are collected in~Table~\ref{tab2}.

If we neglect dynamical corrections ($\gamma = 2$, $\eta = 0$), the MAE leads to a resonance at $4E_a/m^{\text{s}}$ (dashed lines in Fig.~\ref{fig2}(b); $E_a$ is taken from Table~\ref{tab1}).
The downward shift of the resonance energy is due to $\gamma \neq 2$ and an effective $E_a$ smaller than the force theorem estimate (see Table~\ref{tab2}).
$\gamma \neq 2$ is determined by the electronic structure, as discussed in Refs.~\onlinecite{Mills1967,Lounis2010,Lounis2011,Chilian2011}.
The discrepancy in $E_a$ may originate from the following: the value obtained from Eq.~(\ref{mae}) is an energy difference between two orthogonal magnetic orientations, while the LLG model, see Eq.~(\ref{llgeq}), suggests the dynamics depend on the variation of the energy around the equilibrium direction of the adatom spin.
An alternative explanation would be the large DOS peak for Fe and Co at the Fermi energy: the MAE computed from Eq.~(\ref{mae}) is sensitive to peak shifts between the two orientations, which may arise from the frozen potential approximation.
%\footnote{There is a large DOS peak for Fe and Co at the Fermi energy; the MAE computed from Eq.~(\ref{mae}) is sensitive to peak shifts between the two orientations, which may also account for the discrepancy.}.
The damping parameter $\eta$ plays a minor role for Cr and Mn adatoms, but reduces $g$ for Fe and Co by $5\%$ and $22\%$, respectively.
The spin excitation lifetimes are much shorter for the Fe and Co adatoms than for the Cr and Mn adatoms.
This is due to the $\Gamma \propto \omega_{\text{max}}$ scaling of the FWHM, and to the connection between $\eta$ and the slope of $\IM\,\overline{\chi}^{\text{KS}}_{+-}(\omega)$, Fig.~\ref{fig2}(a).
SOC enhances $\eta$ only for the Co adatom, for the same reason (see discussion of Fig.~\ref{fig2}(a) in the main text).

\begin{figure}[t]
  \includegraphics[width=\columnwidth]{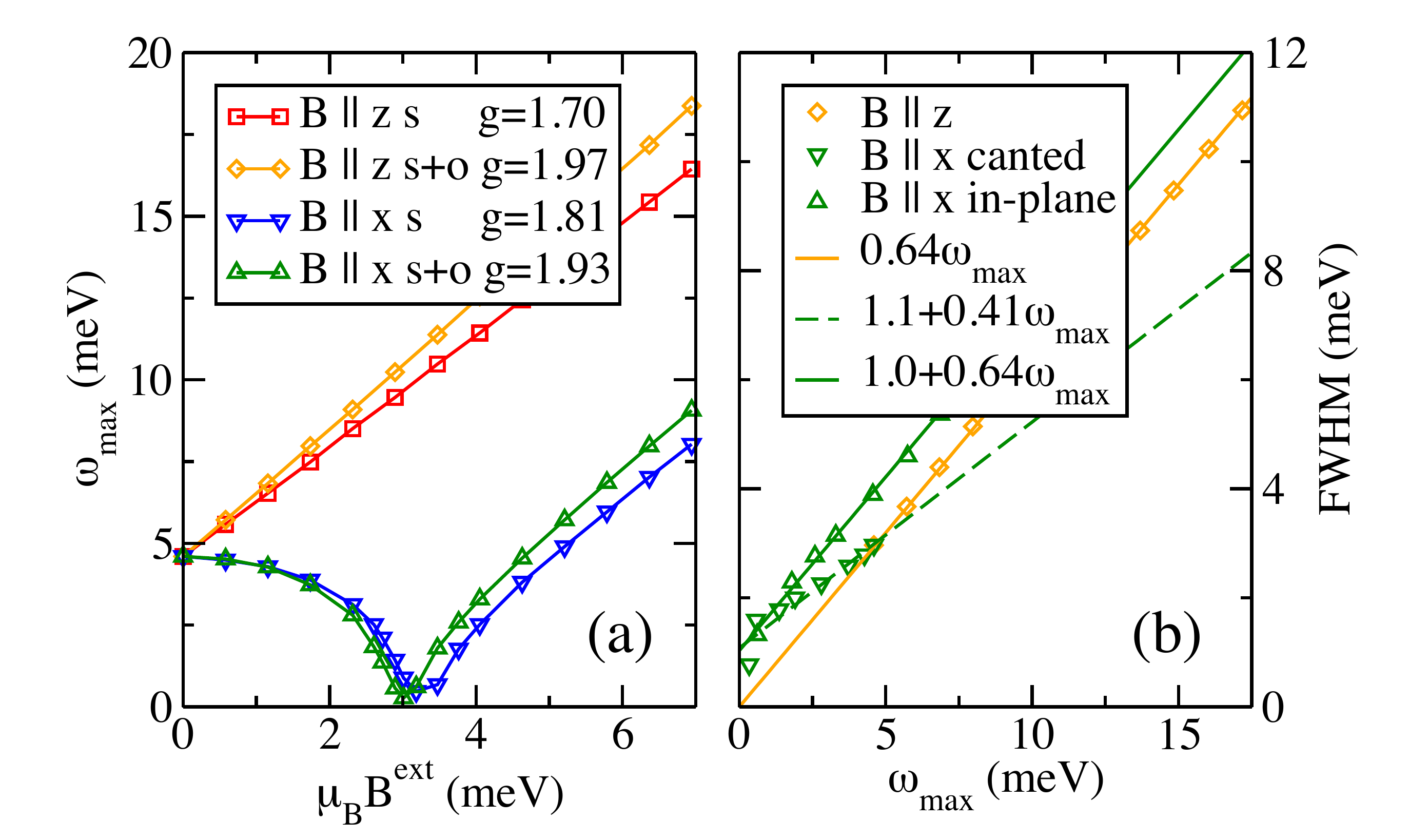}\vspace{-1em}
  \caption{\label{fig3}
  (a) Frequency $\omega_{\text{max}}$ at which $\IM\,\overline{\chi}$ is maximum vs applied external magnetic field, $B^{\text{ext}}$, for the Fe adatom on Cu(111).
  The static field was applied normal to the surface, with spin-only (s, squares) or spin and orbital coupling (s+o, diamonds); and in the surface plane, with spin-only (s, down triangles) or spin and orbital coupling (s+o, up triangles).
  (b) FWHM vs resonance energy, as extracted from the $\IM\,\overline{\chi}$ curves (s+o). %(spin and orbital Zeeman coupling).
  The spin moment is normal to the surface (diamonds), canted (down triangles) or in the surface plane (up triangles).
  Linear fits also shown.}\vspace{-1em}
\end{figure}

\section{Interplay between magnetic anisotropy and magnetic field}\label{sec:feadatom}

Next we focus on the effect of $\vec{B}^{\text{ext}}$ on the spin excitations of the Fe adatom.
First we apply the field parallel to the MAE easy axis, resulting in the linear Zeeman shift. 
$\omega_{\text{max}}$ is shown in Fig.~\ref{fig3}(a) for the spin-only (squares) and spin+orbital (diamonds) Zeeman couplings, see Eq.~(\ref{eq:soczeeman}).
In the spin-only case, $g = 1.70$ ($1.64$ using the LLG model, Table~\ref{tab2}).
This does not depend on SOC and is related to details of the electronic structure~\cite{Chilian2011}. 
When the orbital coupling is included, $g$ rises from 1.70 to 1.97, close to the factor $\big(1 + \frac{m^{\text{o}}}{m^{\text{s}}}\big)$, from Eq.~(\ref{wmax}).
Using $\frac{m^{\text{o}}}{m^{\text{s}}} = 0.17$ and the data in Table~\ref{tab2}, $g = 1.92$ from the LLG model, in good agreement with the first-principles data. 
In Ref.~\onlinecite{Khajetoorians2011}, this system was studied by ISTS.
The measured gap ($\sim\!1$ meV) is lower than the one found in our calculations ($\sim\!4.6$ meV), which may indicate we overestimated the MAE.
The experimental $g$ value ($\sim\!2.1$) is quite close to the one computed including spin and orbital Zeeman couplings ($g = 1.97$).
The linewidth is proportional to the resonance energy, Fig.~\ref{fig3}(b) (diamonds), but the slope is larger in the experimental data.

When the field is applied perpendicular to the MAE easy axis, the equilibrium direction of the Fe spin moment progressively cants away from the surface normal, becoming parallel to $\vec{B}^{\text{ext}}$ (when $\omega_{\text{max}} \approx 0$).
Fig.~\ref{fig3}(a) shows that $\omega_{\text{max}}$ slowly decreases for small $B^{\text{ext}}$, dipping near the critical field, beyond which it increases again, recovering the linear dependence on $B^{\text{ext}}$ for $\omega_{\text{max}} > 5$~meV.
In this regime, $g = 1.81$ (spin Zeeman) or $g = 1.93$ (full Zeeman) differ from the values obtained in the previous linear case.
This arises from the different values of $m^{\text{o}}$, of $\gamma$ and of $\eta$, making the spin dynamics anisotropic.
The FWHM is still linear in $\omega_{\text{max}}$, Fig.~\ref{fig3}(b) (triangles), but it does not extrapolate to zero as in the out-of-plane case.
For the same $\omega_{\text{max}}$, the lifetime of the spin excitation strongly depends on the orientation of the spin moment.
From Fig.~\ref{fig3}(b), with $\omega_{\text{max}} \sim 4.6$~meV, $\tau \approx 110$ fs (out-of-plane) and $\tau \approx 85$ fs (in-plane), which is a 20\% change.

\section{Conclusions}\label{sec:conc}

In this paper we presented a detailed first-principles analysis of the spin dynamics of magnetic adatoms, made possible by an extension of TDDFT to spin-polarized systems including SOC.
We found a novel and invaluable sum rule connecting the spin density to the xc splitting in the presence of SOC.
The key spin dynamics parameters have been extracted from the dynamical magnetic susceptibility, including SOC and spin and orbital Zeeman terms, after mapping to the LLG model, thus legitimating its use down to the atomic limit.
Deviations from standard assumptions in spin dynamics models have been found for the MAE (different from the one computed by the force theorem), the gyromagnetic ratio ($\gamma \neq 2$, which may indicate spin pumping~\cite{Tserkovnyak2002}), and the origin and role of the Gilbert damping, $\eta$ (dominated by Stoner excitations, not by SOC). 
We also find a non-trivial behavior for $g$ and the spin excitation lifetime upon application of a magnetic field along the easy and hard axes of the magnetic anisotropy.
The anisotropic nature of the spin dynamics was established, arising from SOC and $\vec{B}^{\text{ext}}$.

\begin{acknowledgments}
We thank A.~T.~Costa, F.~Guimar\~{a}es and J.~Azpiroz for fruitful discussions.
This work is supported by the HGF-YIG Programme VH-NG-717 (Functional Nanoscale Structure and Probe Simulation Laboratory--Funsilab).
\end{acknowledgments}

\appendix

\section{Derivation of the magnetization sum rule for a non-spin-diagonal Hamiltonian}\label{app:sumrule}

We start from the Kohn-Sham Hamiltonian as given in Eq.~(\ref{eq:soczeeman}) of the main text,
\begin{equation}
  \HH^{\text{KS}}(\vec{r}\,) = \HH^{\text{KS}}_0\sigma_0 + \vec{B}^{\text{xc}}\cdot\vec{\sigma} + \xi\,\vec{L}\cdot\vec{\sigma} + \big(\vec{L} + \vec{\sigma}\big)\cdot\vec{B}^{\text{ext}} \;\;, \label{ksham}
\end{equation}
and the KS Green function, $G(E) = \big(E - \HH^{\text{KS}}\big)^{\text{--}1}$, where the inverse abbreviates the solution of
\begin{equation}
  \sum_{s_1}\big(E\,\delta_{ss_1} - \HH^{\text{KS}}_{ss_1}(\vec{r}\,)\big)\,G_{s_1s'}(\vec{r}\,,\vec{r}\,';E) = \delta(\vec{r}\, - \vec{r}\,')\,\delta_{ss'} \quad.\label{gfdef}
\end{equation}
The summation is over spin components, $s = \{\uparrow, \downarrow\}$.
For a non-local Hamiltonian an integration over the intermediate real-space coordinates would also be present, and the remaining derivation is unaffected.

The $z$-component of the spin density is given by
\begin{equation}
  m_z(\vec{r}\,) = -\frac{1}{\pi}\,\IM\!\int^{E_{\text{F}}}\!\!\!\!\!\!\!\ud E\,\Big(G_{\upup}(\vec{r}\,,\vec{r}\,;E) - G_{\dndn}(\vec{r}\,,\vec{r}\,;E)\Big) \;\;,
\end{equation}
so we must solve for the diagonal (in spatial coordinates and spin labels) parts of the KS GF.

Expanding Eq.~(\ref{gfdef}) we find the following relations among the spin blocks: %(spatial coordinates omitted for brevity):
%\begin{subequations}
%  \begin{eqnarray}
%  \big(E\, - \HH^{\text{KS}}_{\uparrow\uparrow}(\vec{r}\,)\big)\,G_{\uparrow\uparrow}(\vec{r}\,,\vec{r}\,';E) - \HH^{\text{KS}}_{\uparrow\downarrow}(\vec{r}\,)\,G_{\downarrow\uparrow}(\vec{r}\,,\vec{r}\,';E) &=& \delta(\vec{r}\, - \vec{r}\,') \\
%  \big(E\, - \HH^{\text{KS}}_{\downarrow\downarrow}(\vec{r}\,)\big)\,G_{\downarrow\uparrow}(\vec{r}\,,\vec{r}\,';E) - \HH^{\text{KS}}_{\downarrow\uparrow}(\vec{r}\,)\,G_{\uparrow\uparrow}(\vec{r}\,,\vec{r}\,';E) &=& 0 \\
%  \big(E\, - \HH^{\text{KS}}_{\downarrow\downarrow}(\vec{r}\,)\big)\,G_{\downarrow\downarrow}(\vec{r}\,,\vec{r}\,';E) - \HH^{\text{KS}}_{\downarrow\uparrow}(\vec{r}\,)\,G_{\uparrow\downarrow}(\vec{r}\,,\vec{r}\,';E) &=& \delta(\vec{r}\, - \vec{r}\,')  \\
%  \big(E\, - \HH^{\text{KS}}_{\uparrow\uparrow}(\vec{r}\,)\big)\,G_{\uparrow\downarrow}(\vec{r}\,,\vec{r}\,';E) - \HH^{\text{KS}}_{\uparrow\downarrow}(\vec{r}\,)\,G_{\downarrow\downarrow}(\vec{r}\,,\vec{r}\,';E) &=& 0
%  \end{eqnarray}
%\end{subequations}
\begin{subequations}
  \begin{align}
    \big(E\, - \HH^{\text{KS}}_{\uparrow\uparrow}(\vec{r}\,)\big)\,G_{\uparrow\uparrow}(\vec{r}\,,\vec{r}\,';E) &= \delta(\vec{r}\, - \vec{r}\,') \nonumber\\
  \phantom{M} + \HH^{\text{KS}}_{\uparrow\downarrow}&(\vec{r}\,)\,G_{\downarrow\uparrow}(\vec{r}\,,\vec{r}\,';E) \;\;, \\
    \big(E\, - \HH^{\text{KS}}_{\downarrow\downarrow}(\vec{r}\,)\big)\,G_{\downarrow\uparrow}(\vec{r}\,,\vec{r}\,';E) &= 0 \nonumber\\
  \phantom{M} + \HH^{\text{KS}}_{\downarrow\uparrow}&(\vec{r}\,)\,G_{\uparrow\uparrow}(\vec{r}\,,\vec{r}\,';E) \;\;, \\
  \big(E\, - \HH^{\text{KS}}_{\downarrow\downarrow}(\vec{r}\,)\big)\,G_{\downarrow\downarrow}(\vec{r}\,,\vec{r}\,';E) &= \delta(\vec{r}\, - \vec{r}\,') \nonumber\\
  \phantom{M} + \HH^{\text{KS}}_{\downarrow\uparrow}&(\vec{r}\,)\,G_{\uparrow\downarrow}(\vec{r}\,,\vec{r}\,';E) \;\;, \\
  \big(E\, - \HH^{\text{KS}}_{\uparrow\uparrow}(\vec{r}\,)\big)\,G_{\uparrow\downarrow}(\vec{r}\,,\vec{r}\,';E) &= 0 \nonumber\\
  \phantom{M} + \HH^{\text{KS}}_{\uparrow\downarrow}&(\vec{r}\,)\,G_{\downarrow\downarrow}(\vec{r}\,,\vec{r}\,';E) \;\;.
  \end{align}
\end{subequations}

Defining auxiliary GF blocks, $\widetilde G_{\uparrow}(E) = \big(E - \HH^{\text{KS}}_{\uparrow\uparrow}\big)^{\text{--}1}$ and $\widetilde G_{\downarrow}(E) = \big(E - \HH^{\text{KS}}_{\downarrow\downarrow}\big)^{\text{--}1}$, solutions of
\begin{subequations}
  \label{gfaux}
  \begin{eqnarray}
    \big(E\, - \HH^{\text{KS}}_{\uparrow\uparrow}(\vec{r}\,)\big)\,\widetilde G_{\uparrow}(\vec{r}\,,\vec{r}\,';E) &=& \delta(\vec{r}\, - \vec{r}\,') \;\;, \\
    \big(E\, - \HH^{\text{KS}}_{\downarrow\downarrow}(\vec{r}\,)\big)\,\widetilde G_{\downarrow}(\vec{r}\,,\vec{r}\,';E) &=& \delta(\vec{r}\, - \vec{r}\,')\;\;, 
  \end{eqnarray}
\end{subequations}
the previous set of equations can be rewritten as follows:\vspace{-1em}
\begin{subequations}
  \begin{align}
    G_{\uparrow\uparrow}(\vec{r}\,,\vec{r}\,'&;E) = \widetilde G_{\uparrow}(\vec{r}\,,\vec{r}\,';E) \nonumber\\
    + \!\int\!\!\ud\vec{r}_1\,&\widetilde G_{\uparrow}(\vec{r}\,,\vec{r}_1;E)\,\HH^{\text{KS}}_{\uparrow\downarrow}(\vec{r}_1)\,G_{\downarrow\uparrow}(\vec{r}_1,\vec{r}\,';E) \;\;, \label{gupup} \\
    G_{\downarrow\uparrow}(\vec{r}\,,\vec{r}\,'&;E) = 0 \nonumber\\
    + \!\int\!\!\ud\vec{r}_1\,&\widetilde G_{\downarrow}(\vec{r}\,,\vec{r}_1;E)\,\HH^{\text{KS}}_{\downarrow\uparrow}(\vec{r}_1)\,G_{\uparrow\uparrow}(\vec{r}_1,\vec{r}\,';E) \;\;, \label{gdnup} \\
    G_{\downarrow\downarrow}(\vec{r}\,,\vec{r}\,'&;E) = \widetilde G_{\downarrow}(\vec{r}\,,\vec{r}\,';E) \nonumber\\
    + \!\int\!\!\ud\vec{r}_1\,&\widetilde G_{\downarrow}(\vec{r}\,,\vec{r}_1;E)\,\HH^{\text{KS}}_{\downarrow\uparrow}(\vec{r}_1)\,G_{\uparrow\downarrow}(\vec{r}_1,\vec{r}\,';E) \;\;, \label{gdndn} \\
    G_{\uparrow\downarrow}(\vec{r}\,,\vec{r}\,'&;E) = 0 \nonumber\\
    + \!\int\!\!\ud\vec{r}_1\,&\widetilde G_{\uparrow}(\vec{r}\,,\vec{r}_1;E)\,\HH^{\text{KS}}_{\uparrow\downarrow}(\vec{r}_1)\,G_{\downarrow\downarrow}(\vec{r}_1,\vec{r}\,';E) \;\;.\label{gupdn}
  \end{align}
\end{subequations}

After using Eq.~(\ref{gdnup}) and Eq.~(\ref{gupdn}) in Eq.~(\ref{gupup}) and Eq.~(\ref{gdndn}), respectively, we obtain
\begin{subequations}
  \begin{align}
    G_{\uparrow\uparrow}(\vec{r}\,,\vec{r}\,';E) &= \widetilde G_{\uparrow}(\vec{r}\,,\vec{r}\,';E) \nonumber\\
  \phantom{M} + \!\int\!\!\ud\vec{r}_1 &\!\int\!\!\ud\vec{r}_2\,\widetilde G_{\uparrow}(\vec{r}\,,\vec{r}_1;E)\,\HH^{\text{KS}}_{\uparrow\downarrow}(\vec{r}_1)\,\widetilde G_{\downarrow}(\vec{r}_1,\vec{r}_2;E) \nonumber\\\
  &\phantom{MM}\times \HH^{\text{KS}}_{\downarrow\uparrow}(\vec{r}_2)\,G_{\uparrow\uparrow}(\vec{r}_2,\vec{r}\,';E) \;\;, \\
  G_{\downarrow\downarrow}(\vec{r}\,,\vec{r}\,';E) &= \widetilde G_{\downarrow}(\vec{r}\,,\vec{r}\,';E) \nonumber\\
  \phantom{M} + \!\int\!\!\ud\vec{r}_1 &\!\int\!\!\ud\vec{r}_2\,\widetilde G_{\downarrow}(\vec{r}\,,\vec{r}_1;E)\,\HH^{\text{KS}}_{\downarrow\uparrow}(\vec{r}_1)\,\widetilde G_{\uparrow}(\vec{r}_1,\vec{r}_2;E) \nonumber\\
  &\phantom{MM}\times \HH^{\text{KS}}_{\uparrow\downarrow}(\vec{r}_2)\,G_{\downarrow\downarrow}(\vec{r}_2,\vec{r}\,';E) \;\;,
  \end{align}
\end{subequations}
%
%Highlighting the spin structure of the GF and introducing the auxiliary Green function blocks,
%\begin{equation}
%G(E) = \begin{pmatrix}
%E - \HH^{\text{KS}}_{\uparrow\uparrow} & - \HH^{\text{KS}}_{\uparrow\downarrow} \\
%- \HH^{\text{KS}}_{\downarrow\uparrow} & E - \HH^{\text{KS}}_{\downarrow\downarrow}
%\end{pmatrix}^{-1}
%\!\!= \begin{pmatrix} G_{\uparrow\uparrow}(E) & G_{\uparrow\downarrow}(E) \\ G_{\downarrow\uparrow}(E) & G_{\downarrow\downarrow}(E) \end{pmatrix} \;\;,\quad
%\widetilde G(E) = \begin{pmatrix}
%E - \HH^{\text{KS}}_{\uparrow\uparrow} & 0 \\
%0 & E - \HH^{\text{KS}}_{\downarrow\downarrow}
%\end{pmatrix}^{-1}
%\!\!= \begin{pmatrix} \widetilde G_\uparrow(E) & 0 \\ 0 & \widetilde G_\downarrow(E) \end{pmatrix} \;\;,
%\end{equation}
%the desired components of the KS GF can be expressed as
and the solution of these two equations can be represented in an abbreviated form as
\begin{subequations}
  \begin{align}
    G_{\uparrow\uparrow}(E) &= \Big(E - \HH^{\text{KS}}_{\uparrow\uparrow} - \HH^{\text{KS}}_{\uparrow\downarrow}\,\widetilde G_\downarrow(E)\,\HH^{\text{KS}}_{\downarrow\uparrow}\Big)^{-1} \;\;, \\
    G_{\downarrow\downarrow}(E) &= \Big(E - \HH^{\text{KS}}_{\downarrow\downarrow} - \HH^{\text{KS}}_{\downarrow\uparrow}\,\widetilde G_\uparrow(E)\,\HH^{\text{KS}}_{\uparrow\downarrow}\Big)^{-1} \;\;.
  \end{align}
\end{subequations}
Here the inverse operation is for the $(\vec{r}\,,\vec{r}\,')$ dependence, as in Eq.~(\ref{gfaux}); the spin dependence was already solved for.

By the following operator identity
\begin{equation}
  A^{-1} - B^{-1} = A^{-1}(B - A)\,B^{-1} = B^{-1}(B - A)\,A^{-1} \;\;, \label{eq:inviden}
\end{equation}
the difference between the two GF blocks of interest is given by
\begin{align}
  &G_{\uparrow\uparrow}(\vec{r}\,,\vec{r}\,';E) - G_{\downarrow\downarrow}(\vec{r}\,,\vec{r}\,';E) \nonumber\\
&= \!\int\!\!\ud\vec{r}_1\!\int\!\!\ud\vec{r}_2\,G_{\uparrow\uparrow}(\vec{r}\,,\vec{r}_1;E)\,\Delta(\vec{r}_1,\vec{r}_2;E)\,G_{\downarrow\downarrow}(\vec{r}_2,\vec{r}\,';E) \nonumber\\
&= \!\int\!\!\ud\vec{r}_1\!\int\!\!\ud\vec{r}_2\,G_{\downarrow\downarrow}(\vec{r}\,,\vec{r}_1;E)\,\Delta(\vec{r}_1,\vec{r}_2;E)\,G_{\uparrow\uparrow}(\vec{r}_2,\vec{r}\,';E) \;\;,
\end{align}
with the energy dependent splitting defined by
\begin{align}
  \Delta(\vec{r}\,,\vec{r}\,';E) &= \Big(\HH^{\text{KS}}_{\uparrow\uparrow}(\vec{r}\,) - \HH^{\text{KS}}_{\downarrow\downarrow}(\vec{r}\,)\Big)\,\delta(\vec{r}\,-\vec{r}\,') \nonumber\\
& + \HH^{\text{KS}}_{\uparrow\downarrow}(\vec{r}\,)\,\widetilde G_\downarrow(\vec{r}\,,\vec{r}\,';E)\,\HH^{\text{KS}}_{\downarrow\uparrow}(\vec{r}\,') \nonumber\\
&- \HH^{\text{KS}}_{\downarrow\uparrow}(\vec{r}\,)\,\widetilde G_\uparrow(\vec{r}\,,\vec{r}\,';E)\,\HH^{\text{KS}}_{\uparrow\downarrow}(\vec{r}\,') \;\;.
\end{align}
Now the concrete form of the splitting is given, making use of the terms in Eq.~(\ref{ksham}).
This is especially meaningful if the xc field points along the $z$ direction and is responsible for most of the spin splitting:
\begin{align}
  &\Delta(\vec{r}\,,\vec{r}\,';E) = 2\,\Big(B^{\text{xc}}(\vec{r}\,) \!+ \xi(r)\,L_z + B_z^{\text{ext}}(\vec{r}\,)\Big)\,\delta(\vec{r} - \vec{r}\,') \nonumber\\
&+ \Big(\xi(r) L_-\!+\!B_-^{\text{ext}}(\vec{r}\,)\Big) \widetilde G_\downarrow(\vec{r}\,,\vec{r}\,';E) \Big(\xi(r') L_+\!+\!B_+^{\text{ext}}(\vec{r}\,')\Big) \nonumber\\
&- \Big(\xi(r) L_+\!+\!B_+^{\text{ext}}(\vec{r}\,)\Big) \widetilde G_\uparrow(\vec{r}\,,\vec{r}\,';E) \Big(\xi(r') L_-\!+\!B_-^{\text{ext}}(\vec{r}\,')\Big) \;\;,
\end{align}
with the combinations $L_\pm = L_x \pm \iu Ly$ and $B_\pm^{\text{ext}} = B_x^{\text{ext}} \pm \iu B_y^{\text{ext}}$.
Defining the spin flip KS susceptibilities as
\begin{align}
  \chi_{+-}^{\text{KS}} &= \frac{1}{4}\Big(\chi_{xx}^{\text{KS}} - \iu\chi_{xy}^{\text{KS}} + \iu\chi_{yx}^{\text{KS}} + \chi_{yy}^{\text{KS}}\Big) \;\;, \\%\quad\text{and}\quad
  \chi_{-+}^{\text{KS}} &= \frac{1}{4}\Big(\chi_{xx}^{\text{KS}} + \iu\chi_{xy}^{\text{KS}} - \iu\chi_{yx}^{\text{KS}} + \chi_{yy}^{\text{KS}}\Big) \;\;,
\end{align}
from the definition in terms of GFs and Pauli matrices
%\begin{equation}
%  \chi^{\text{KS}}_{\alpha\beta}(\vec{r}\,,\vec{r}\,';\omega) = -\frac{1}{\pi}\!\int^{E_{\text{F}}}\!\!\!\!\!\!\!\ud E\;
%  \Tr \Big[ \sigma_\alpha\,G(\vec{r}\,,\vec{r}\,';E + \omega + \iu0)\,\sigma_\beta\,\IM\,G(\vec{r}\,',\vec{r}\,;E)
%  + \sigma_\alpha\,\IM\,G(\vec{r}\,,\vec{r}\,';E)\,\sigma_\beta\,G(\vec{r}\,',\vec{r}\,;E-\omega-\iu0) \Big] \;\;, \label{kssuscgf}
%\end{equation}
\begin{align}
  \chi^{\text{KS}}_{\alpha\beta}(\omega) = -\frac{1}{\pi}\!\int^{E_{\text{F}}}\!\!\!\!\!\!\!\ud E\;
  &\Tr \Big[ \sigma_\alpha\,G(E + \omega + \iu0)\,\sigma_\beta\,\IM\,G(E) \nonumber\\
  &+ \sigma_\alpha\,\IM\,G(E)\,\sigma_\beta\,G(E-\omega-\iu0) \Big] \;\;, %\label{kssuscgf}
\end{align}
we arrive at the magnetization sum rule,
\begin{align}
  m_z(\vec{r}\,) &= 2\!\int\!\!\ud\vec{r}\,'\,\chi^{\text{KS}}_{+-}(\vec{r}\,,\vec{r}\,';0)\,B^{\text{xc}}(\vec{r}\,') + \Delta m_z(\vec{r}\,) \nonumber\\
&= 2\!\int\!\!\ud\vec{r}\,'\,\chi^{\text{KS}}_{-+}(\vec{r}\,,\vec{r}\,';0)\,B^{\text{xc}}(\vec{r}\,') + \Delta m_z(\vec{r}\,) \;\;,
\end{align}
where the two equivalent variants follow from the two forms of the operator identity, Eq.~(\ref{eq:inviden}), and the small correction $\Delta m_z(\vec{r}\,)$ arises from the spin-orbit coupling and external magnetic field contributions to $\Delta(\vec{r}\,,\vec{r}\,';E)$, (\emph{i.e.}~excluding the xc part).

\section{Kohn-Sham susceptibility in a basis derived from KKR scattering solutions}\label{app:suscbasis}

In the atomic sphere approximation (ASA), the KKR GF has the form~\cite{Papanikolaou2002}:
%\begin{subequations}
\begin{equation}
  G_{ij}^{s}(\vec{r}\,,\vec{r}\,';E) = \sum_{LL'} Y_L(\hat{r})\,G_{iL,jL'}^{s}(r,r';E)\,Y_{L'}(\hat{r}') %\vspace{-1em}
\end{equation}
with
\begin{align}\label{eq:gfils}
  G_{iL,jL'}^{s}(r,r';E) &= R_{i\ell}^s(r;E)\,G_{iL,jL'}^s(E)\,R_{j\ell}^s(r';E) \nonumber\\
  &+ \delta_{ij}\,\delta_{LL'} R_{i\ell}^s(r_<;E)\,H_{i\ell}^s(r_>;E)  \;\;.
\end{align}
%\end{subequations}
The position arguments of the GF are measured from the nearest atomic site, labelled $i$.
The orientation is denoted $\hat{r}$ and the length $r$; furthermore, $r_< = \min\{r,r'\}$ and $r_> = \max\{r,r'\}$.
The angular dependence is expanded in spherical harmonics $Y_L(\hat{r})$, with composite index $L=(\ell,m)$, and the two spin components are labelled by $s$.
The radial functions $R_{i\ell}^s(r;E)$ and $H_{i\ell}^s(r;E)$ are solutions of the Schr\"odinger equation for the KS potential centered in site $i$ for a given energy $E$, which are regular and irregular at the nuclear position, respectively.
The structural GF, $G_{iL,jL'}^s(E)$, contains the information about the geometrical arrangement of the atomic sites and the multiple scattering contributions.
We assumed a collinear magnetic state and no SOC, for simplicity of presentation.

We use a basis of radial functions constructed from normalized regular scattering solutions computed at several energies $E_b$ within the range of the valence states (four energy values are sufficient for the basis construction):
\begin{equation}
  \phi_{i\ell b}^s(r) = \frac{R_{i\ell}^s(r;E_b)}{\int_0^{R} \ud r\,r^2\,R_{i\ell}^s(r;E_b)^2} \quad. %\vspace{1em}
\end{equation}
For fixed $i$ and $\ell$, by orthogonalizing the overlap matrix
\begin{equation}
  \mathcal{O}^{i\ell}_{bb'} = \int_0^{R}\!\!\ud r\,r^2\,\phi_{i\ell b}^s(r)\,\phi_{i\ell b'}^s(r)
\end{equation}
and keeping only the two largest eigenvalues, we form two linear combinations of the reference basis functions by using the respective eigenvectors, which become the basis functions for atom $i$ and angular momentum channel $\ell$.

The most general form of the KS GF (with SOC and non-collinear magnetism), in the KKR representation and in our chosen basis, is thus
\begin{equation}
  G_{iL,jL'}^{ss'}(r,r';E) = \sum_{bb'} \phi_{i\ell b}^s(r)\,G_{iLb,jL'b'}^{ss'}(E)\,\phi_{j\ell' b'}^{s'}(r') \quad.
\end{equation}

The KS susceptibility, Eq.~(\ref{kssusc}), is then also naturally expressed in this basis,
\begin{widetext}
  \begin{equation}
    \chi^{\text{KS}}_{\alpha\beta}(\vec{r}\,,\vec{r}\,';\omega)
    = \!\!\!\sum_{L_1s_1b_1\cdots}\!\!\!Y_{L_1}(\hat{r}) Y_{L_2}(\hat{r}) \phi_{i\ell_1 b_1}^{s_1}\!(r) \phi_{i\ell_2 b_2}^{s_2}\!(r)
    \,\chi^{\text{KS},\,s_1s_2s_3s_4}_{\alpha\beta,\,iL_1L_2b_1b_2,\,jL_3L_4b_3b_4}(\omega)
    \,\phi_{j\ell_3 b_3}^{s_3}\!(r') \phi_{j\ell_4 b_4}^{s_4}\!(r') Y_{L_3}(\hat{r}') Y_{L_4}(\hat{r}') \quad.
  \end{equation}
\end{widetext}

The size of the matrices involved is as follows.
For the GF, the number of rows or columns is $N_{a} \times N_L \times N_s \times N_b = 1 \times (\ell_{\text{max}}+1)^2 \times 2 \times 2 = 64$ when taking only an adatom into account ($N_a=1$), with $\ell_{\text{max}} = 3$ and two radial basis functions.
For the KS susceptibility matrix, the number of rows or columns is $N_\alpha \times N_{a} \times \big( N_L \times N_s \times N_b \big)^2 = 4 \times 64^2 = 16384$, as we have four Pauli matrices ($\alpha = x, y, z, 0$), with the same assumptions as for the previous example.

\section{Dynamical magnetic susceptibility from the Landau-Lifshitz-Gilbert equation}\label{app:llg}

The aim is to linearize the LLG equation,
\begin{equation}
  \frac{\ud\vec{m}}{\ud t} = -\gamma\,\vec{m} \times \vec{B}^{\text{eff}} + \eta\,\frac{\vec{m}}{m}\!\times\!\frac{\ud\vec{m}}{\ud t}
\end{equation}
with the effective field
\begin{equation}
  \vec{B}^{\text{eff}} = -\frac{\partial E}{\partial\vec{m}} = B^{\text{eff}}\,\hat{e}_z + b_x(t)\,\hat{e}_x + b_y(t)\,\hat{e}_y
\end{equation}
where $B^{\text{eff}}\,\hat{e}_z$ is the static part, and $b_x(t)\,\hat{e}_x + b_y(t)\,\hat{e}_y$ is the small time-dependent transverse part.
Under these assumptions, the same applies to the magnetization,
\begin{equation}
  \vec{m}(t) = m_x(t)\,\hat{e}_x + m_y(t)\,\hat{e}_y + M_z\,\hat{e}_z \;\;,
\end{equation}
with $m_x, m_y \ll M$.

After Fourier transforming, $\frac{\ud}{\ud t} \rightarrow -\iu\omega$, the dynamical transverse magnetic susceptibility can be extracted from the LLG equation as
\begin{equation}
  \chi_{+-}(\omega) = \frac{M_z\omega_0}{2B^{\text{eff}}}\,\frac{(1+\eta^2)\omega_0 - \omega + \iu\eta\omega}{(\omega - \omega_0)^2 + (\eta\omega_0)^2}
  \;\;,\quad \omega_0 = \frac{\gamma B^{\text{eff}}}{1 + \eta^2} \;\;. \label{eq:llgimchi}
\end{equation}
This form can be used to fit the first-principles TDDFT data and extract all the parameters.

The resonance peak location is obtained from
\begin{align}
  \frac{\ud}{\ud\omega}\IM\,\chi_{+-} &= 0 \nonumber\\
  &\Longrightarrow\;\; (\omega - \omega_0)^2 + (\eta\omega_0)^2 - 2\omega(\omega - \omega_0) = 0 \nonumber\\
  &\Longrightarrow\;\; \omega_{\text{max}} = \frac{\gamma B^{\text{eff}}}{\sqrt{1 + \eta^2}} \;\;,
\end{align}
which is the result quoted in the main text.

\bibliography{bibliography}

\end{document}